\documentclass[amsmath, amssymb, aps, prl, reprint, preprintnumbers, superscriptaddress, nofootinbib, nobibnotes]{revtex4-2}
\usepackage{graphicx}
\usepackage{lineno}
\usepackage{colortbl}
\usepackage{siunitx}
\usepackage{dcolumn}
\usepackage{blindtext}
\usepackage[caption=false]{subfig}
\usepackage[table, usenames, dvipsnames]{xcolor}
\usepackage{array}
\newcolumntype{L}[1]{>{\raggedright\let\newline\\\arraybackslash\hspace{0pt}}m{#1}}
\newcolumntype{C}[1]{>{\centering\let\newline\\\arraybackslash\hspace{0pt}}m{#1}}
\newcolumntype{R}[1]{>{\raggedleft\let\newline\\\arraybackslash\hspace{0pt}}m{#1}}

\newcommand{\triumf}{\affiliation{TRIUMF, 4004 Wesbrook Mall, Vancouver, British Columbia V6T 2A3, Canada}}
\newcommand{\nd}{\affiliation{Department of Physics, University of Notre Dame, Notre Dame, Indiana 46556, USA}}

\newcommand{\losalamos}{\affiliation{Theoretical Division, Los Alamos National Laboratory, Los Alamos, New Mexico 87545, USA}}
\newcommand{\cta}{\affiliation{Center for Theoretical Astrophysics, Los Alamos National Laboratory, Los Alamos, New Mexico 87545, USA}}
\newcommand{\ncstate}{\affiliation{Department of Physics, North Carolina State University, Raleigh, North Carolina 27695, USA}}

\newcommand\tstrut{\rule{0pt}{3ex}}         

\begin{document}
\title{The need for a local nuclear physics feature in the neutron-rich rare-earths \\ to explain solar $r$-process abundances}

\author{Nicole~Vassh}
\email{nvassh@triumf.ca, nvassh@nd.edu}\triumf\nd
\author{Gail~C.~McLaughlin}\ncstate
\author{Matthew~R.~Mumpower}\losalamos\cta
\author{Rebecca~Surman}\nd 
\date{\today}

\begin{abstract}
We apply Markov Chain Monte Carlo to predict the masses required to form the observed solar $r$-process rare-earth abundance peak. Given highly distinct astrophysical outflows and nuclear inputs, we find that results are most sensitive to the $r$-process dynamics (i.e. overall competition between reactions and decays), with similar mass trends predicted given similar dynamics. We show that regardless of whether fission deposits into the rare-earths or not, our algorithm consistently predicts the need for a local nuclear physics feature of enhanced stability in the neutron-rich lanthanides.
\end{abstract}

\maketitle

For more than 60 years the solar abundances have been providing clues to the astrophysical origins of heavy elements \cite{B2FH}. Although the era of multi-messenger astronomy presents new paths to understanding single events \cite{AbbottGW170817,Cowperthwaite2017,Villar,CoteGW170817,Zimmerman2020,Capano+2020,Cfpaper}, the solar abundances still serve as the key informant of the contributions of a given site to the enrichment of the Solar System. To model the dominant astrophysical source of rapid neutron capture ($r$-process) elements in a modern way, statistical methods offer a fresh and innovative approach. With such methods, observational and experimental data can be used to trace back to more fundamental nuclear physics properties \cite{UtamaPiekarewiczMass,Neufcourt+drip,SangalinePratt,UtamaPiekarewiczCrust,WuMacFadyen,Drischler+2020}. 

The rare-earth abundance peak seen in the $r$-process residuals at $A\sim164$ is ideal for first applications of such statistical methods to the solar abundances since many nuclear species of importance have yet to be probed experimentally but some relevant nuclear physics information is available to guide the calculation. Unlike the second ($A\sim130$) and third ($A\sim195$) peaks linked to the neutron shell closures at $N=82$ and $N=126$ respectively, the mechanism by which the rare-earth peak forms is presently uncertain. Since the abundances in this region are highly sensitive to the astrophysical environment in which heavy element synthesis occurs, the answer to rare-earth peak origins can provide hints to the source of $r$-process elements in our galaxy. Rare-earth abundances are also highly sensitive to the nuclear properties of lanthanides thereby permitting small changes to inputs introduced via statistical methods to greatly influence the predicted outcome. Such an investigation is timely both due to the importance of lanthanide abundances for kilonova signals as well as the significant advancements at current and upcoming nuclear physics facilities.

It is possible to use an MCMC procedure to derive mass adjustments to the Duflo-Zuker (DZ) mass model \cite{DufloZuker} which, when used in the r-process, produce consistency with the rare earth solar data \cite{REMM1,REMM2,OrfordVassh2018,VasshMCMC}. Using MCMC with astrophysical outflows typical of simulations of accretion disk wind ejecta \cite{Surman+08,Metzger+2008,Perego+14,Fernandez+15}, predicts two different peak formation mechanisms that are associated with two broad classes of reaction dynamics (i.e overall competition between reactions and decays) \cite{VasshMCMC}. In outflows with hot dynamics, (n,$\gamma$)$\rightleftarrows$($\gamma$,n) equilibrium persists for long timescales and shapes the $r$-process path (location of most abundance species at a given $Z$). In contrast, in cold outflows photodissociation falls out of equilibrium early with the competition between neutron capture and $\beta$-decay largely determining how the $r$-process proceeds. When compared to independent, precision mass measurements, MCMC mass surface predictions in the hot case are more consistent than those in the cold case \cite{VasshMCMC}. The primary impactful feature in the mass surface for the hot case is at  $N=104$ and produces a persistent `pile-up' at this neutron number with the path having its highest abundances at $N=104$ for many proton numbers \cite{OrfordVassh2018,VasshMCMC}. This feature is just outside the latest measurements at $N=104$.

\begin{table*}[ht!]
\caption{Description of all astrophysical outflows and nuclear inputs considered in this work. Note the initial conditions ($Y_e$, entropy, and density) are all reported at 8 GK with the entropies given assuming the SFHo equation of state \cite{SFHo2013}.}
\label{tab:trajs}
\centering
\begin{tabular}{L{2.5cm}L{5.0cm}L{0.75cm}L{0.75cm}L{1.25cm}L{1.25cm}L{1.25cm}L{3.95cm}}
\toprule
\tstrut
Label  & Description & Type & $Y_e$ & {Initial Entropy (s/$k_B$)} & {Initial Density (g/cm$^3$)} & Fission Yields & Fission Rates \\ 
\hline
\hline
\tstrut
$hot\_wind$ & parameterized, low entropy outflow as from an accretion disk & hot & 0.2 & 30 & 8.4$\times10^6$ & N/A & N/A\\
\hline
$cold\_wind$ & parameterized, low entropy outflow as from an accretion disk & cold & 0.2 & 10 & 6.1$\times10^7$ & N/A & N/A\\
\hline
$hot\_dyn\_sf$ & NSM dynamical ejecta simulation \cite{MendozaTemis} with reheating & hot & 0.01 & 9 & 1.1$\times10^9$ & 50/50 & instantaneous spontaneous fission for $A\ge250$\\
\hline
$hot\_dyn\_5050$ & NSM dynamical ejecta simulation \cite{MendozaTemis} with reheating & hot & 0.01 & 9 & 1.1$\times10^9$  & 50/50 & neutron-induced, $\beta$-delayed, and spontaneous fission from FRLDM barriers + DZ masses\\
\hline
$hot\_dyn\_GEF16$ & NSM dynamical ejecta simulation \cite{MendozaTemis} with reheating & hot & 0.01 & 9 & 1.1$\times10^9$  & GEF16 & same as $hot\_dyn\_5050$ \\
\hline
$hot\_dyn\_GEF18$ & NSM dynamical ejecta simulation \cite{MendozaTemis} with reheating & hot & 0.01 & 9 & 1.1$\times10^9$  & GEF18 & same as $hot\_dyn\_5050$ \\
\hline
$cold\_dyn\_sf$ & NSM dynamical ejecta simulation \cite{GorielyJanka}, no reheating & cold & 0.01 & 22 & 2.8$\times10^7$ & 50/50 & instantaneous spontaneous fission for $A\ge250$\\
\hline
$cold\_dyn\_5050$ & NSM dynamical ejecta simulation \cite{Rosswog} with reheating & cold & 0.02 & 4 & 7.0$\times10^{10}$ & 50/50 & neutron-induced, $\beta$-delayed, and spontaneous fission from FRLDM barriers + DZ masses\\
\hline
$cold\_dyn\_GEF16$ & NSM dynamical ejecta simulation \cite{Rosswog} with reheating & cold & 0.02 & 4 & 7.0$\times10^{10}$ & GEF16 & same as $cold\_dyn\_5050$\\
\hline
$cold\_dyn\_GEF18$ & NSM dynamical ejecta simulation \cite{Rosswog} with reheating & cold & 0.02 & 4 & 7.0$\times10^{10}$ & GEF18 & same as $cold\_dyn\_5050$\\
\botrule
\end{tabular}
\end{table*} 

These conclusions were reached for moderately neutron-rich conditions, leaving a process which could play a key role in shaping rare-earth abundances remained unexplored, that is, fission. One might expect fission product deposition into the rare earth region to impact peak formation. In this case the rare-earth abundances could be influenced by the properties of actinide species lying far away in the nuclear chart rather than being purely shaped by the local nuclear features of lanthanide isotopes. In fact there has been speculation that late-time fission deposition could be  the exclusive origin of the rare-earth peak \cite{GorielySPY}.  Here we consider the sensitivity of the predictions of local nuclear physics feature to fission deposition by using yield predictions from the 2016 and 2018 versions of the GEF code \cite{GEF} (GEF16 and GEF18), as well as a symmetric split of all fissioning nuclei in half (50/50) and compare to results obtained in moderately neutron rich conditions.
 
To this end, we consider very neutron-rich outflows ($Y_e\sim0.01$) that significantly populate fissioning nuclei.
Such outflows have been consistently predicted to dominate the dynamical ejecta from a neutron star merger \cite{Rosswog,GorielyJanka,Bovard2017,Radice18,Foucart2020}. In addition to the participation of fission, further distinctions between very neutron-rich conditions as compared to moderately neutron-rich cases exist. First, the $r$-process path tends to reach isotopes further from stability. Second, the higher neutron richness implies a greater potential for late-time neutron capture to play a role in shaping the rare-earth peak. Thus, considering ejecta which reaches fissioning nuclei applies our method to cases which have the maximal possible deviation from the moderately neutron-rich cases considered in \cite{VasshMCMC} due to both the distinct evolution of their hydrodynamic properties as well as the impact of fission deposition on abundances. Therefore this work reports on the masses predicted to be needed for rare-earth peak formation given the largest set of astrophysical and theoretical nuclear physics inputs considered to date.

\begin{figure*}
    \centering
    \includegraphics[width=17cm]{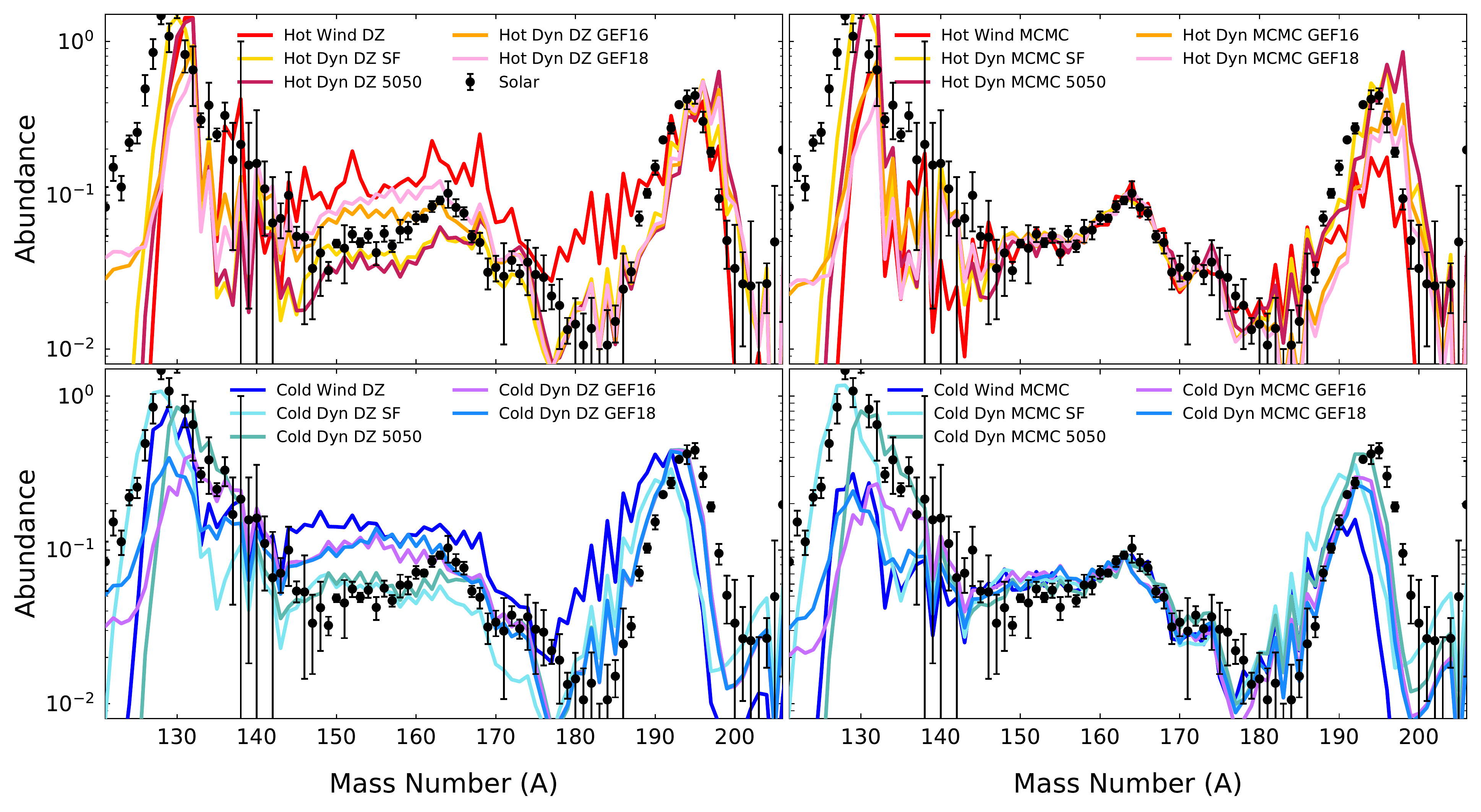}%
   \caption{(Top left) Baseline abundances for all hot astrophysical outflows and fission treatments outlined in Table~\ref{tab:trajs} (with $\chi^2$ values of 200, 121, 236, 212, and 211 respectively). (Top right) Final abundances for the hot conditions given the mass solutions found for each case using our MCMC procedure (with $\chi^2$ values of 23, 37, 40, 36, and 35). (Bottom left) Same as top left but for all cold astrophysical outflows considered (with $\chi^2$ values of 286, 361, 211, 499, and 488). (Bottom right) Same as top right but for the case of cold conditions (with $\chi^2$ values of 22, 106, 68, 185, and 148). On the left abundances are the unscaled network output with mass fractions summing to one whereas on the right results are scaled to the solar abundances between $A=150-180$. The solar abundances and uncertainties were derived from those in \cite{Arnould07,Goriely99} as described in \cite{VasshMCMC}.}
\label{fig:allabund}%
\end{figure*} 

As was done with the moderately neutron-rich cases, we consider several very neutron-rich conditions with a range of hydrodynamic properties whose initial conditions are outlined in Table~\ref{tab:trajs}. In addition to diverse initial conditions, the cases considered also have distinct temperature/density evolutions. For instance, $0.1$ sec after the 8 GK initial conditions, the densities and temperatures of the moderately neutron-rich cases $hot\_wind$ and $cold\_wind$ have dropped to $\sim10^5,10^3$ g/cm$^3$ and $3.2,0.7$ GK respectively while the densities and temperatures of the very neutron-rich dynamical ejecta outflows $hot\_dyn\_sf$, $cold\_dyn\_sf$, and $cold\_dyn\_5050$ have fallen to $\sim10^6,10^4,10^4$ g/cm$^3$ and $0.5,1.0,0.4$ GK respectively (with the low temperature, low density condition being that of tidal tail ejecta and the highest density condition being that of shock-heated ejecta). 

In very neutron-rich ejecta, the nuclear reheating effects on the temperature become more pronounced than in moderately neutron-rich conditions. We therefore consider the impact of reheating on the dynamics of the outflow by using a cold condition with a heating efficiency of zero ($cold\_dyn\_sf$) as well as outflows for which the reheating efficiency is non-zero. We note that even with reheating included, there remain outflow conditions under which the dynamics remain of the cold type ($cold\_dyn\_5050$, $cold\_dyn\_GEF16$, $cold\_dyn\_GEF18$). Considering a range of outflow evolutions is an important part of analyzing the robustness of our previous findings that cases with distinct dynamics (hot vs. cold) require distinct lanthanide masses in order to form the rare-earth peak. Thus for the present investigation of very neutron-rich ejecta, we again separately consider peak formation in hot and cold scenarios.
 
We show abundance patterns for all the astrophysical outflows and nuclear inputs outlined in Table~\ref{tab:trajs} in Figure~\ref{fig:allabund}. In all cases, our baseline abundances are flat on average with no peak structure or show a rough peak which is off center from that seen the solar data. Here the influence of the fission prescription on the final abundances is well demonstrated since prior to applying our MCMC procedure, cases with 50/50 yields ($hot\_dyn\_sf$, $hot\_dyn\_5050$, $cold\_dyn\_sf$, and $cold\_dyn\_5050$) concentrate deposition near the second peak and leave the rare-earth abundances to be structured by local effects whereas in the case of GEF16 yields ($hot\_dyn\_GEF16$ and $cold\_dyn\_GEF16$) deposition occurs between the second peak and rare-earth peak. A stronger influence from fission deposition with GEF18 yields ($hot\_dyn\_GEF18$ and $cold\_dyn\_GEF18$) is evident from the enhanced abundances near $A\sim160$. When considering the $\chi^2$ fit in the $A=150-180$ region, this enhancement in abundances from fission deposition to the left of the rare-earth peak produces a higher $\chi^2$ value when the abundances scaled to the peak are compared to the solar data.

In the right panel we show the results of the application of the MCMC method. We start with a mass parameterization:
\begin{equation}
M(Z,N) = M_{DZ}(Z,N)+a_{N}e^{-(Z-C)^2/2f}
\end{equation}
where $M_{DZ}(Z,N)$ is the DZ mass of nuclear species $(Z,N)$, $a_{N}$ are the mass adjustments determined by the MCMC procedure, and the exponential acts to center the adjustments in the neutron-rich region where masses are largely unmeasured.  At each MCMC step, neutron capture rates and beta decay properties are adjusted according to the test masses, and a nucleosynthesis calculation is performed and compared to solar data to set the evolution of the Markov chain. Our general procedure was first introduced in \cite{REMM1,REMM2} and revised and refined in \cite{OrfordVassh2018,VasshMCMC}.  Even though here we consider cases with higher initial $\chi^2$ values than were considered in \cite{VasshMCMC}, in all cases we are able to obtain substantially lower $\chi^2$ values after applying our MCMC procedure, as can be seen in the right panels of Figure~\ref{fig:allabund}. For instance, for all the hot, very neutron-rich dynamical ejecta considered here, our MCMC procedure is able to find solutions with $\chi^2\le40$, with most being around $35$, which is comparable to what the MCMC was able to achieve when faced with moderately neutron-rich conditions. Therefore, despite a wide variety of astrophysical outflow types and nuclear inputs, our MCMC algorithm is able to find a solution for the masses of neutron-rich rare-earths which gives abundances that are significantly more consistent with the solar data.

\begin{figure}
    \centering
    \includegraphics[width=8.85cm]{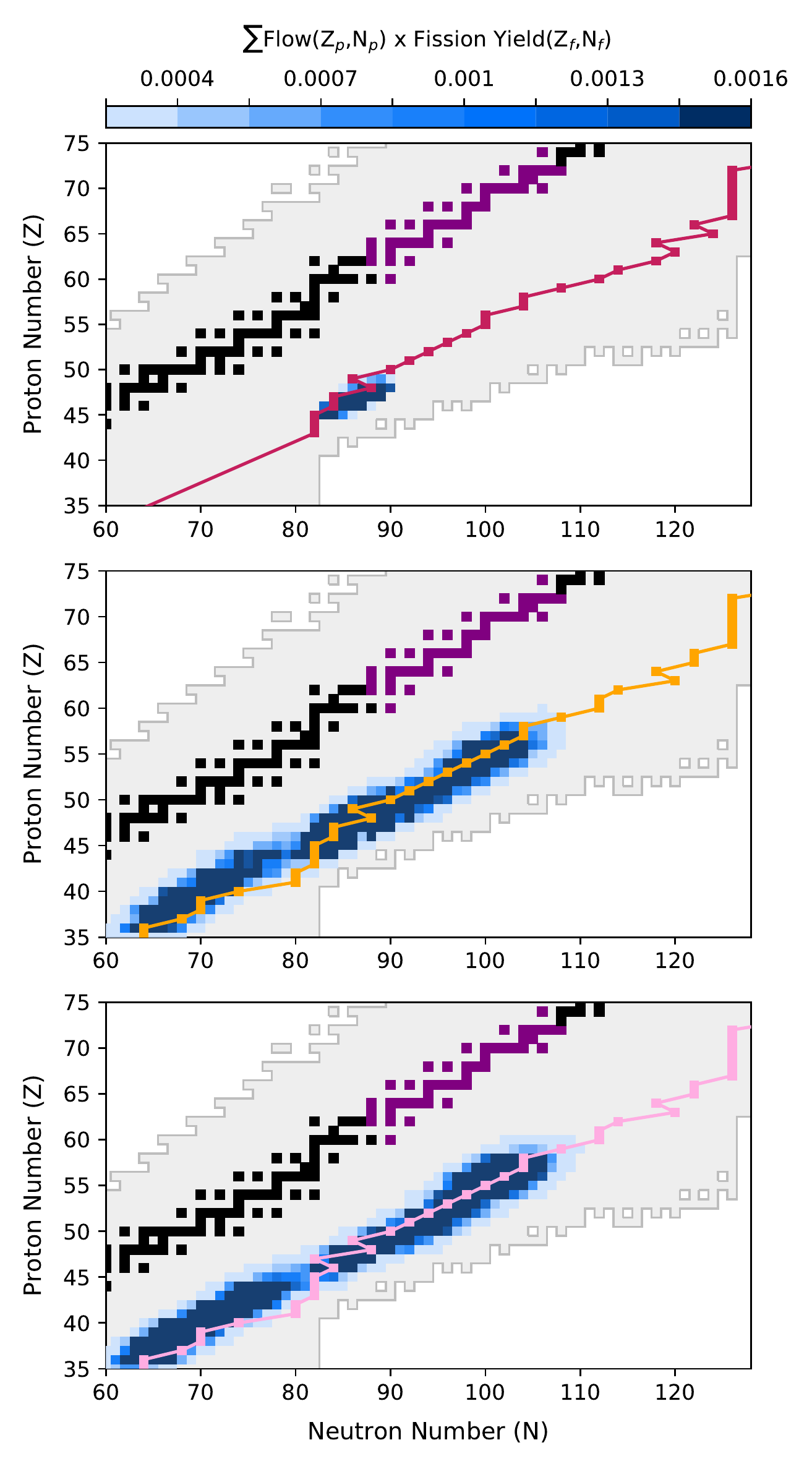}%
   \caption{The summed fission flow (rate$\times$abundance) of neutron-induced and $\beta$-delayed fission for each fissioning species multiplied by its fission yield to demonstrate deposition with (top) 50/50 splits, (middle) GEF16 yields, and (bottom) GEF18 yields (cases $hot\_dyn\_5050$, $hot\_dyn\_GEF16$ and $hot\_dyn\_GEF18$, respectively) at a given instance in time (1.4 seconds with temperature 1.1 GK and density 3.4$\times10^3$g/cm$^3$). For reference, the dark pink (top), orange (middle) and light pink (bottom) show the $r$-process path at this time and the grey shows the DZ dripline. Purple boxes highlight the stable nuclei with $A=150-180$, whose abundances following the $\beta$-decay of $r$-process species determines the structure of the rare-earth peak, and black boxes denote all other stable species.} 
\label{fig:yielddep}%
\end{figure}

We next more explicitly demonstrate the influence of fission deposition on rare-earth peak abundances using the hot dynamical ejecta cases. 
In Figure \ref{fig:yielddep} we show a snapshot of the summed neutron-induced and $\beta$-delayed fission flow of fissioning nuclei multiplied by their fission yield for the three distinct fission yield treatments considered in this work. The $r$-process path is also shown to demonstrate that fission deposition is occurring during a time at which the nuclei which will go on to form the rare-earth peak are undergoing pile-up. Given a symmetric 50/50 split for all nuclei, deposition remains isolated near the $N=82$ shell closure. The yields predicted by the GEF16 model transition from asymmetric to symmetric across the nuclear chart (see \cite{VasshFiss}), which populates neutron-rich isotopes that will decay back to set abundances on the left side of the rare-earth peak. When compared to GEF16, the yields predicted by GEF18 have enhanced asymmetries for several nuclei, such as those near $N=184$, thereby having greater amounts of deposition into regions set to populate central species in the rare-earth peak. Note that in an astrophysical outflow with fission cycling such as those considered here, nuclei in the rare-earths go through a wave or multiple waves of first being maximally populated to then be depleted as they undergo capture to heavier species, with fission eventually repopulating these nuclei so that the process can then repeat. Since this all occurs as the ejecta is expanding and cooling, there is a time sensitive connection between fission product deposition and the ultimate population of lanthanides versus actinides. Despite this complex interplay between the lanthanides and actinides, any isotopes that remain in the rare-earth region after the conditions can no longer support capture up to the actinides must decay back to stability and be subject to local nuclear structure influences.  

\begin{figure*}
    \centering
    \includegraphics[width=8.95cm]{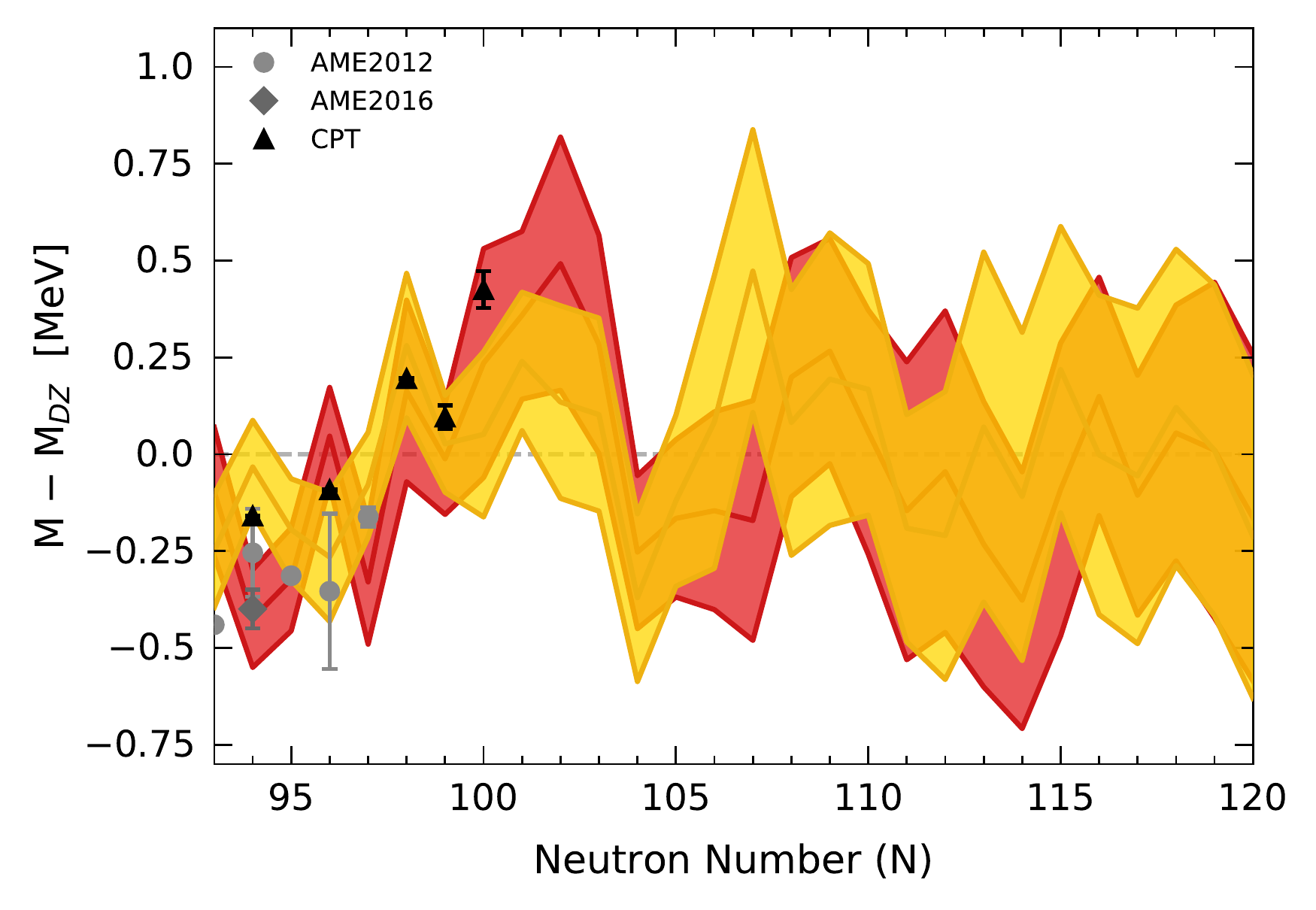}%
    \includegraphics[width=8.95cm]{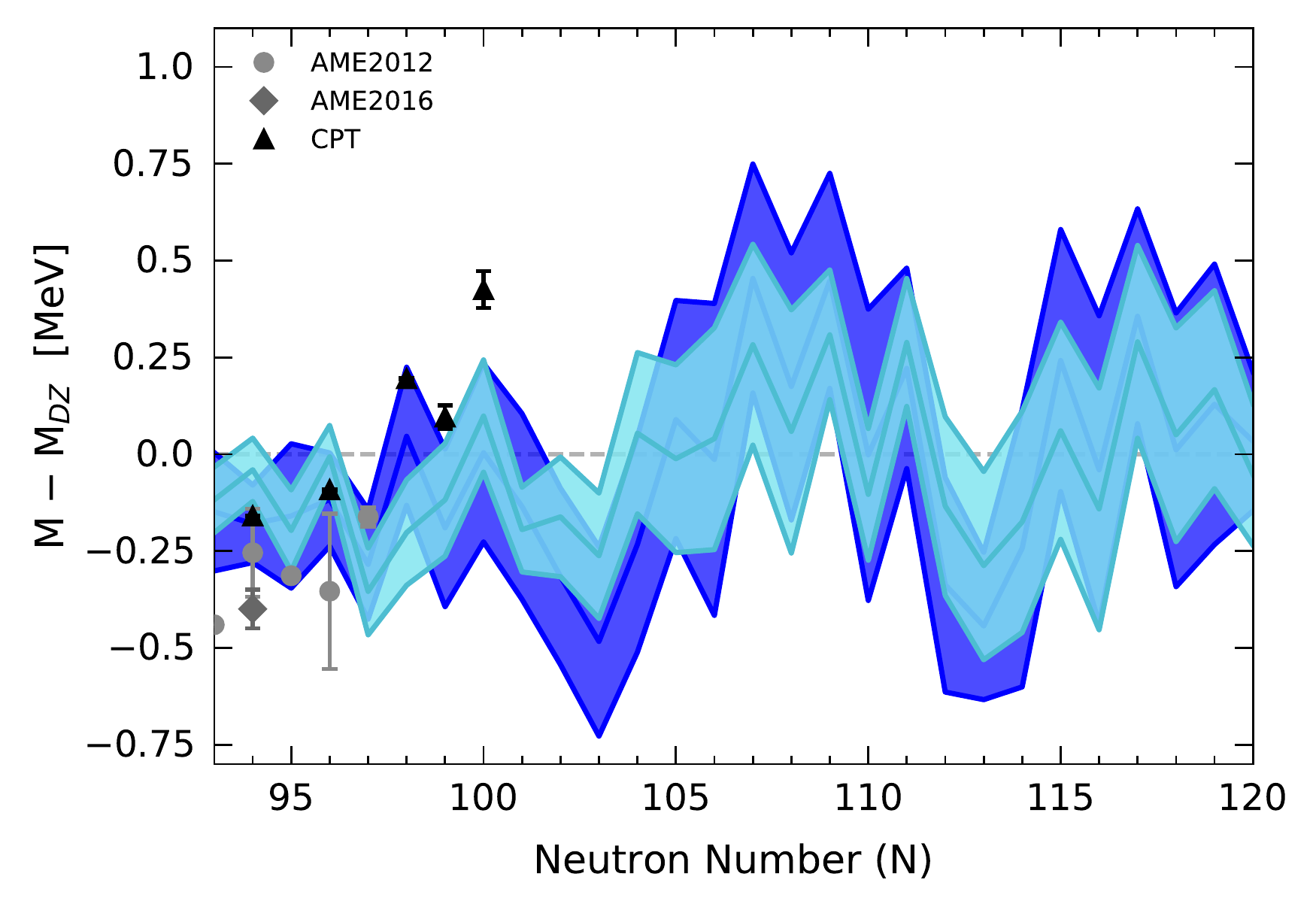}%
    \hspace{0.5cm}
    \includegraphics[width=8.95cm]{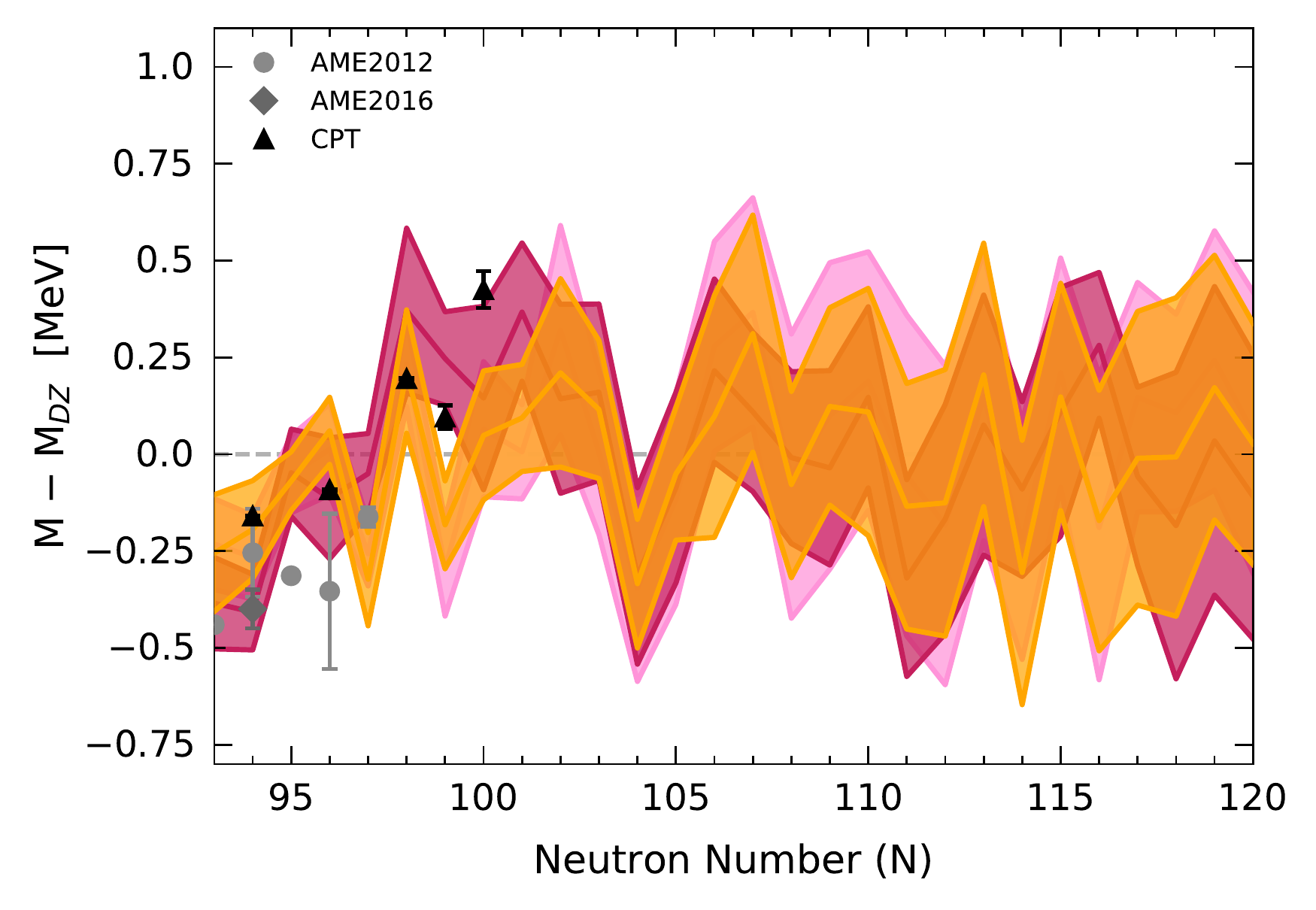}%
    \includegraphics[width=8.95cm]{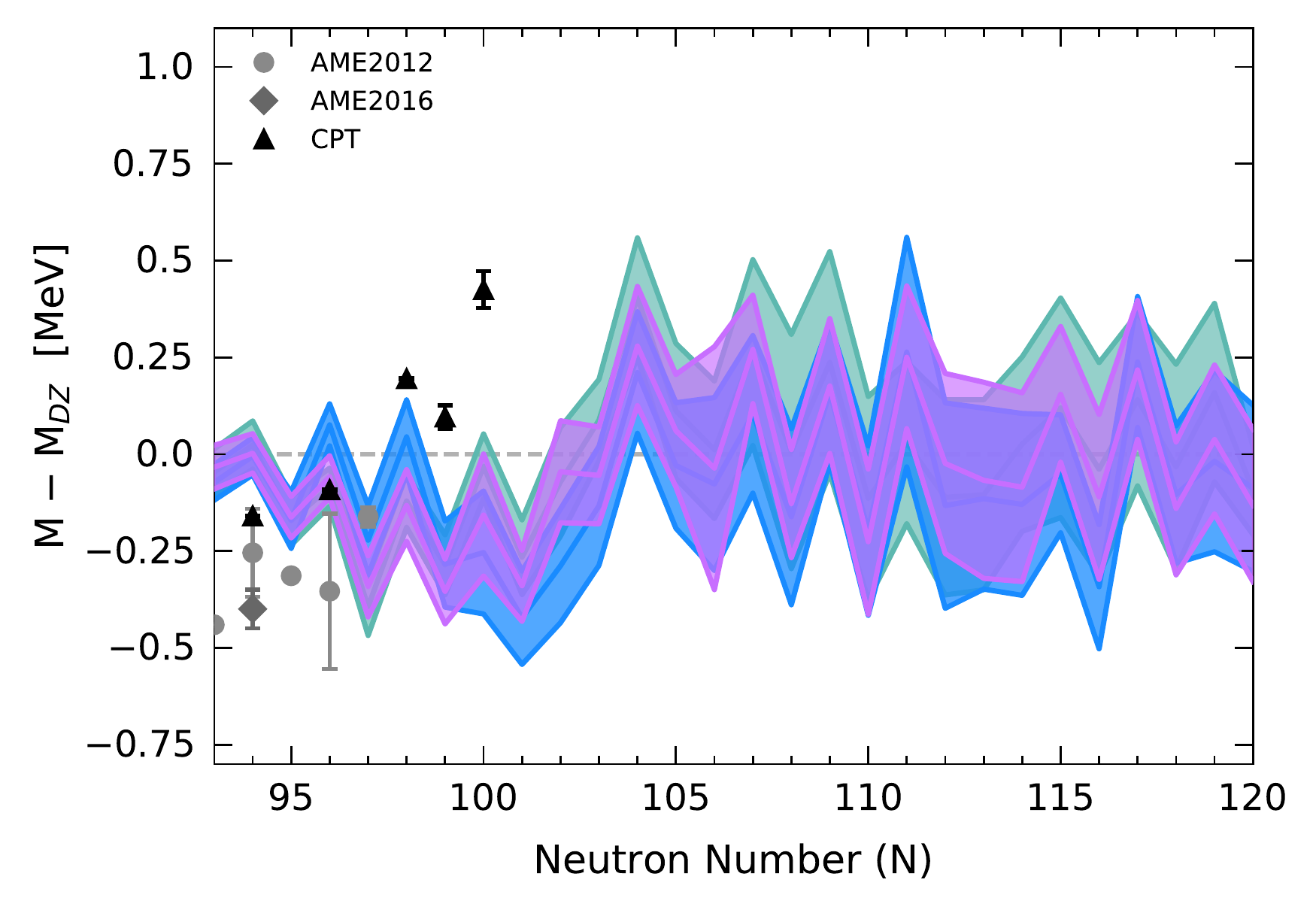}%
   \caption{The MCMC predicted masses for neodymium ($Z=60$), relative to the DZ mass model, given (top left) the moderately neutron-rich $hot\_wind$ outflow explored in \cite{VasshMCMC} (red band) compared to results with the very neutron-rich condition $hot\_dyn\_sf$ (yellow band),  (top right) the moderately neutron-rich condition $cold\_wind$ explored in \cite{VasshMCMC} (dark blue band) compared to results for the very neutron-rich condition $cold\_dyn\_sf$ (light blue band). (Bottom left) Results for the very neutron-rich conditions of $hot\_dyn\_5050$ (dark pink band) compared to cases $hot\_dyn\_GEF16$ (orange band) and $hot\_dyn\_GEF18$ (light pink band) which apply yields that give some late time deposition into the rare-earth region. (Bottom right) Results given the very neutron-rich conditions of $cold\_dyn\_5050$ (teal band) as compared to the $cold\_dyn\_GEF16$ (purple band) and $cold\_dyn\_GEF16$ (medium blue band) cases. The AME2012 data \citep{AME2012} used to guide the calculation is also shown, along with AME2016 \citep{AME2016} and CPT at CARIBU \citep{OrfordVassh2018} data of which the calculation was not informed.}
\label{fig:allmasssurf}%
\end{figure*}

The masses derived using our MCMC method in all hot and cold dynamical ejecta cases are shown in Fig.~\ref{fig:allmasssurf}, with an explicit comparison to our previous findings for moderately neutron-rich ejecta in the top panels. For the very neutron-rich outflow results in the top panels, the hot dynamical ejecta case $hot\_dyn\_sf$ uses an outflow tracer from \cite{MendozaTemis} which accounts for nuclear reheating. The cold dynamical ejecta case $cold\_dyn\_sf$ uses an outflow tracer from \cite{GorielyJanka} and exemplifies an extreme of cold conditions since the efficiency of nuclear heating is assumed to be zero. For the cases featured in the top panels, we consider a simplified fission treatment in which all nuclei with $A\ge250$ spontaneously fission at a very fast rate to then deposit near the second $r$-process peak and not in the rare-earth peak by applying symmetric 50/50 fission yields. Such a simplification was first considered for two main reasons. Firstly it allows to build predictions with comparable statistics to the moderately neutron-rich cases (both derived from 50 MCMC runs) since running the nucleosynthesis network at each MCMC step is much less costly than is the case when more proper fission rates and yields are considered. Secondly our simplified fission treatment removes the effects of late time fission deposition on rare-earth abundances (by contributing exclusively to the abundances near $A\sim130$) thereby permitting us to isolate how our predictions change due solely to the distinctions in the outflow properties of very neutron-rich cases as compared to moderately neutron-rich cases. We note that as in \cite{VasshMCMC}, we center our mass adjustments at $C=60$ in the hot cases and $C=58$ in the cold cases (the $r$-process paths in cold cases tend to lie further from stability, therefore in test runs mass adjustments near $C=58$ were favored). As can be seen from the top panels of Fig.~\ref{fig:allmasssurf}, although some differences in the masses predictions exist, the MCMC results for the very neutron-rich $hot\_dyn\_sf$ case have key similarities to the results when the moderately neutron-rich case $hot\_wind$ is considered. Both show a rise in the mass surface near $N=102$ followed by the dip at $N=104$ to be the features primarily responsible for the peak structure near $A=164$. Key similarities are also observed in results for the cold outflows $cold\_wind$ and $cold\_dyn\_sf$ except that an influence from a nuclear physics feature at $N=110$ emerges in the dynamical ejecta case considered here, whereas in the moderately neutron-rich $cold\_wind$ case $N=103$ coupled with $N=108$ features were sufficient to form the peak. Therefore although the neutron-richness does influence the details of peak formation, similar peak formation mechanisms are needed for conditions with similar $r$-process dynamics. Particularly, in the case of hot outflows, the mechanism which we previously found to be responsible for peak formation, that is the pile-up at $N=104$, is robust.

We now discuss whether the fission treatment significantly modifies the expected local nuclear masses of the lanthanides needed in order to form the peak. To do so, we move away from the simplified treatment previously described and apply a more proper treatment of fission rates by utilizing predictions from CoH and BeoH Hauser-Feshbach codes \cite{CoH} when DZ masses and FRLDM barriers \cite{FRLDM} are assumed (similar to the treatment described in \cite{VasshFiss}). We also advance the fission yield treatment by implementing the three models previously described. The bottom panels of Fig.~\ref{fig:allmasssurf} show the MCMC predictions when these nuclear data inputs are considered in both hot dynamical ejecta ($hot\_dyn\_5050$, $hot\_dyn\_GEF16$, and $hot\_dyn\_GEF18$) and cold dynamical ejecta ($cold\_dyn\_5050$, $cold\_dyn\_GEF16$, and $cold\_dyn\_GEF18$). Additionally, we note that here for the cold case we consider a merger tidal tail ejecta tracer from \cite{Rosswog} where we have included the effect of nuclear reheating on the trajectory but this cases nevertheless retains its cold dynamics. In all scenarios considered in the bottom panel, running the nucleosynthesis network on each MCMC step is costly due to the need to follow hundreds of fission products. Therefore to reduce the computational cost, we apply the method outlined in \cite{VasshFiss} whereby only the fissioning nuclei which most influence the abundances are treated with GEF16 and GEF18 yields and a 50/50 split is applied for all other species. Although this reduces the cost of running the network significantly, there is nevertheless a greater computational expense than in the simplified fission case considered in the top panels of Fig.~\ref{fig:allmasssurf}. Therefore since here our primary aim is to test the robustness of the similarities we see in the predictions for hot versus cold cases as well as the robustness of the nuclear physics feature we repeatedly see at $N=104$ in hot cases, our MCMC predictions for calculations which apply a more proper fission treatment are sufficiently explored through 25 MCMC runs in the hot dynamical ejecta cases and 15 MCMC runs in the cold dynamical ejecta cases. 

The influence of the fission yield treatment is evident in the bottom panels of Fig.~\ref{fig:allmasssurf} for both the hot and cold cases. For hot dynamical ejecta, the build-up in the abundances to the left of the rare-earth peak due to the deposition from GEF16 and GEF18 fission yields (as can be seen in Fig.~\ref{fig:allabund}) must be suppressed. This is achieved by introducing a dip feature in the mass surface at $N=99$ which was not needed for the $hot\_dyn\_5050$ and $hot\_dyn\_sf$ cases for which fission deposition plays no role in setting rare-earth abundances. In the case of $hot\_dyn\_GEF18$, the GEF18 yield prescription also requires a mass surface feature at $N=108$ in order for the late-time deposition seen with this model to stay contained in the rare-earth region, although this works to mostly fill in the right side of the peak. Strikingly, and most importantly, in the hot case all yield models considered point to the now familiar need for a nuclear physics feature of enhanced stability at $N=104$ to produce the pile-up which ultimately creates the rare-earth peak at $A=164$. For the cold dynamical ejecta MCMC runs, it is primarily a feature at $N=101$ which is responsible for peak formation which is in tension with the latest precision mass measurements. Therefore despite the distinct ways in which nuclei are populated in the rare-earth region, our MCMC algorithm finds that all fission yield models and all astrophysical outflows considered require local assistance to form the peak via a nuclear physics feature emerging in neutron-rich lanthanides.

In this work, we have presented the MCMC mass predictions which are capable of forming the $r$-process rare-earth abundance peak given highly distinct astrophysical outflows and nuclear data treatments. While our study of moderately neutron-rich nuclei in \cite{VasshMCMC} was able to show that: (1) outflows with distinct reaction dynamics (hot vs. cold) require distinct mass surfaces in order to form the rare-earth peak and (2) the mass predictions in hot outflows are most consistent with the latest measurements up to $N=102$, the investigation presented in this work is able to significantly broaden what can be concluded from our MCMC procedure. The detailed exploration of peak formation in very neutron-rich ejecta presented here shows that such conditions require mass trends which are similar to those found given moderately neutron-rich outflows. This highlights that it is the $r$-process dynamics (e.g. hot vs. cold), rather than the exact details of the hydrodynamic properties, which is most influential on peak formation. This work therefore could be used alongside simulation advancements to infer the astrophysical site at which solar lanthanides where dominantly formed by considering the degree of participation of photodissociation in simulation outflows. Additionally the calculations presented in this work show that although the fission treatment does produce some noteworthy differences in the masses predictions, the algorithm nevertheless consistently predicts the need for a local nuclear physics feature in the neutron-rich lanthanides in order to form the peak, despite the differences in fission deposition and outflow properties considered here. In all results given astrophysical outflows with cold dynamics, our algorithm predicts features which are in tension with the latest experimental data, although here features at neutron numbers which  are several neutrons higher than what has been probed are important. In the case of hot outflows, our MCMC mass predictions persistently allude to the presence of a nuclear physics feature causing enhanced stability in the lanthanides at $N=104$, just a few neutron numbers outside current measurements. Therefore, since our algorithm finds that fission deposition must be aided by local lanthanide nuclear features to form the peak, near future theoretical and experimental campaigns to map out neutron-rich lanthanide properties, as could be possible at ARIEL at TRIUMF, FRIB, and N=126 Factory at ANL, are well poised to explicitly address a long-standing mystery of heavy element production, that is, the origin the rare-earth peak.

\section*{Acknowledgments}
The work of N.V., G.C.M., M.R.M., and R.S. was partly supported by the Fission In R-process Elements (FIRE) topical collaboration in nuclear theory, funded by the U.S. Department of Energy. Additional support was provided by the U.S. Department of Energy through contract numbers DE-FG02-02ER41216 (G.C.M), DE-FG02-95-ER40934 (R.S.), and DE-SC0018232 (SciDAC TEAMS collaboration, R.S.). R.S. and G.C.M also acknowledge support by the National Science Foundation N3AS Hub Grant No. PHY-1630782 and Physics Frontiers Center No. PHY-2020275. M.R.M. was supported by the US Department of Energy through the Los Alamos National Laboratory. Los Alamos National Laboratory is operated by Triad National Security, LLC, for the National Nuclear Security Administration of U.S.\ Department of Energy (Contract No.\ 89233218CNA000001). 
This work was partially enabled by the National Science Foundation under Grant No. PHY-1430152 (JINA Center for the Evolution of the Elements). This work utilized the computational resources of the Laboratory Computing Resource Center at Argonne National Laboratory (ANL LCRC) and the University of Notre Dame Center for Research Computing (ND CRC). We specifically acknowledge the assistance of Stanislav Sergienko (ANL LCRC) and Scott Hampton (ND CRC). This manuscript has been released via Los Alamos National Laboratory report number LA-UR-XX-XXXXX.

\bibliography{REMMrefs2}
\bibliographystyle{apsrev4-1}

\end{document}